\documentclass[aps,prl,showpacs,amsmath,amssymb,twocolumn]{revtex4}
\usepackage{graphicx}
\usepackage{epsf}
\begin{document}
\title{Superconducting Vortices and Elliptical Ferromagnetic Textures.}
\author{M. Amin Kayali}
\affiliation{Department of Physics, Texas A \& M University
College Station, Texas 77843-4242, USA.}

\begin{abstract}
In this article an analytical and numerical study of
superconducting thin film with ferromagnetic textures of
elliptical geometries in close proximity is presented. The
screening currents induced in the superconductor due to the
magnetic texture are calculated. Close to the
superconducting transition temperature $T_c$ the spontaneous
creation of superconducting vortices becomes energy favorable
depending on the value of the magnetization and the geometrical
quantities of the magnetic texture. The creation of vortices by elliptic dots
is more energy favorable than those created by circular ones. The 
superconductor covered by elliptic dots array exhibits anisotropic transport 
properties.

\end{abstract}
\noindent
\maketitle
The interaction between superconducting vortices and ferromagnetic
textures in heterogeneous ferromagnetic-superconducting systems
has been under intensive study in the past few years. The
static and dynamical phases of these systems have shown richness
and promise for superconductor based technology. Basically, such a system 
consists of a superconducting thin film placed either on the
top or under a ferromagnet and a layer of insulator oxide is sandwiched
between them to suppress proximity effects and spin diffusion. Different
settings and geometries for both the superconductor and ferromagnet were
studied and analyzed both experimentally and theoretically 
\cite{schull}-\cite{KP}. The interaction between a circular FM dot and 
superconducting thin film is well studied for magnetization
distributions either parallel or perpendicular to the SC layer.
Pinning effects and transport in superconductors interacting with periodic 
array of ferromagnetic dots \cite{mosch} or periodic stripe domains structure 
\cite{erdin}, \cite{KP} were studied.\\ \indent
These studies showed that in some range of temperatures and above a threshold 
value of the dot magnetization, the interaction between superconductivity and 
ferromagnetism dominates over other interactions in the system. At such 
conditions for the temperature and magnetization a spontaneous 
vortex phase forms in the superconductors. Additionally, experimental 
measurements and theoretical predictions assert that the usage of a periodic 
pinning array of ferromagnetic textures results in higher values of the 
superconducting critical current than those obtained from random pinning by 
lattice inhomogenities. Almost all previous studies treated the interaction 
between superconducting vortices and FM textures focused on textures with 
circular geometry. However, many studies have investigated the interaction 
of SC vortices with elliptical lattice inhomogeneities such as elliptic 
holes and elliptic columnar defects \cite{buzdin1}, \cite{buzdin2}, 
\cite{buzdin4}, \cite{buzdin5}. To our knowledge the 
interaction between SC vortices and elliptic ferromagnetic dots (EMD) has 
not yet been studied. 

In this article, I present a theoretical treatment of the interaction 
between SC vortices and elliptic ferromagnetic textures. The study of the 
interaction between elliptic dots and superconductivity is interesting since 
its results when the dot's eccentricity $\mathcal{E}$ is zero correspond 
to those known results for circular dots. Another interesting limit is when 
$\mathcal{E}\rightarrow 1$ which mimic a system of long magnetic stripe 
domains interacting with an SC film. This article is organized such that in 
the first section, we calculate the magnetic fields and screening currents 
for a system of elliptic FM dot on the top of an SC thin film. Section two 
is devoted to calculate the total energy and pinning forces. In section 
three, we discuss the dependence of the energy on the eccentricity of the 
FM texture. More detailed analysis of pinning forces and qualitative 
discussion of thetransport in these systems will be presented. Concluding 
remarks and a summary of this work will be given in the last section.

\begin{figure}[t]
  \centering
  \includegraphics[angle=0,width=3.0in,totalheight=2.0in]{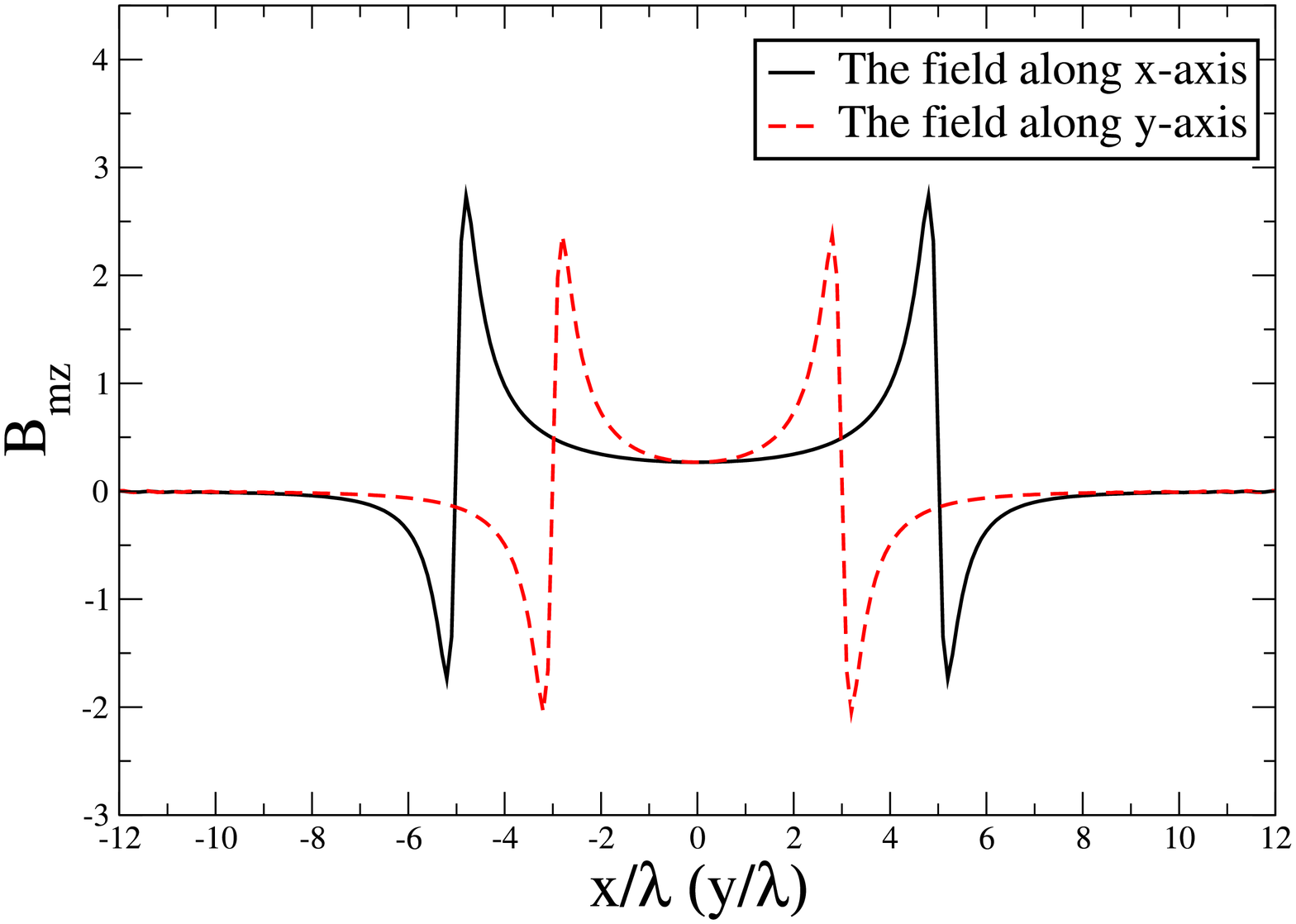}
  \caption{The magnetic field in units of $\frac{2 m_0}{\lambda}$ measured
    along the semi major axis (solid line) and semi-minor axis (dashed line) 
    for EMD with $R_1 =5$ and $R_2 =3$}
  \label{fig1}
\end{figure}

Since the SC and FM are electronically separated, the interaction between 
them is mediated via their magnetic fields. The FM dot produces a 
magnetic field which penetrate the SC film and alters the distribution of its
screening current. In turns the SC generates a magnetic field in 
and out of its plane which interact with the FM dot. The problem of finding the
 magnetic field and screening currents must be solved self consistently. To 
do so, let us consider a superconducting thin film of thickness $d_s$,
whose coherence length is $\xi$ and its penetration depth is $\lambda$ in the
$xy$-plane. We place on the top of it at a distance $D\ll \lambda$ an
elliptical ferromagnetic dot of major axis $R_1$ and minor axis $R_2$.
Let the dot magnetization $\bf{M}$ be directed along the $z$-axis,
the magnetization distribution can be written as
\begin{eqnarray}
{\bf{M}}(x,y,z)=m_0 \Theta (1 -\frac{x^2}{R_1^2}-\frac{y^2}{R_2^2}) 
\delta (z-D) \hat{z}
\end{eqnarray}
\noindent
where $m_0$ is the 2D magnetization, $\Theta (r)$ is the  step function and
$\delta (r)$ is Dirac delta function . In the presence of 
the superconductor the magnetic vector potential $\bf{A}_m$ of the 
dot satisfy the London-Maxwell equation
\begin{eqnarray}
\bf{\nabla}\times \bf{\nabla} \times \bf{A}_m +\frac{1}{\lambda} \bf{A}_m
\delta (z)= 4 \pi \bf{\nabla}\times \bf{M}
\label{eqn2}
\end{eqnarray}
\noindent
Accepting the gauge $\bf{\nabla}\cdot\bf{A}_m =0$, and using the 
integral Fourier representation for ${\bf{A}}_m$, we finds  
\begin{eqnarray}
\tilde{\bf{A}}_m ({\bf{K}}) &=& \frac{-8\pi^2 \imath m_0 R_1 R_2 J_1 
\left(G(k_x,k_y)\right)}
{G(k_x,k_y)\left(k_z^2 +q^2\right)} \times \nonumber \\
& &\left(e^{\imath k_z D} -
\frac{e^{-q D}}{1+2\lambda q} \right) \hat{z} \times {\bf{q}}
\label{eqn7}
\end{eqnarray}
where $\tilde{{\bf{A}}}_m$ is the magnetic dot vector potential in 
Fourier representation and ${\bf{q}}=k_x \hat{x}+k_y \hat{y}$ is 
Fourier wave vector in the plane of the SC. The function $J_n(r)$ is the 
$n$-th order Bessel Function, and $G(k_x,k_y)=\sqrt{R_1^2 k_x^2 
+R_2^2 k_y^2}$. By using $\bf{B}=\bf{\nabla}\times\bf{A}$, the components 
of the dot's magnetic field can be calculated
\begin{eqnarray}
B_{mz} &=& m_0 R_1 R_2 \int  
\frac{q J_1(G(k_x,k_y))Z(k_x,k_y)}{G(k_x,k_y)}\times \nonumber \\
& &e^{-\imath(k_x x+k_y y)} d^2 q
\label{eqn9}
\end{eqnarray}
\begin{eqnarray}
B_{mj} &=& \imath m_0 R_1 R_2 \int \frac{k_j J_1 (G(k_x,k_y))
W(k_x,k_y)}{G(k_x,k_y)}\times \nonumber \\
& &e^{-\imath (k_x x +k_y y)}d^2 q 
\end{eqnarray}
where $j=x,y$, while $Z(k_x,k_y)=e^{-q|z-D|} -\frac{e^{-q(|z|+D)}}{1 
+2\lambda q}$, and $W(k_x,k_y)=e^{-q|z-D|} sign(z-D)- 
\frac{e^{-q(|z|+D)}}{1 +2\lambda q}sign(z)$. 
The in-plane components of the EMD magnetic fields have a jump 
at $z=0$ which should be taken into account. The $z$-component 
of the dot's magnetic field is depicted in Fig.(\ref{fig1}). 
The magnetic field of the dot changes strongly across the dot's circumference  
due to large values of ${\bf{\nabla}}.{\bf{M}}$ there. If vortices are 
present in the SC film then the total magnetic field is a linear 
superposition of the field from the EMD and that of the vortices. The 
z-component of the magnetic field due to a singly quantized SV centered at 
the origin \cite{abrik} reads
\begin{eqnarray}
B_v^z (x,y,z)=\frac{\phi_0}{2\pi}\int_0^\infty \frac{q
J_0(q\sqrt{x^2 +y^2}) e^{-q|z|}}{1 +2\lambda q} dq 
\label{eqnvor}
\end{eqnarray}
where $\phi_0=\frac{\pi\hbar c}{e}$ is the magnetic flux quantum.


Let us assume that there are an $N$ spontaneously created vortex in the 
superconductor. The total energy of for a system of $N$ vortices coupled to  
an FM texture is made up of four different contributions and can be written as
\begin{eqnarray}
U=U_{sv} +U_{vv} +U_{mv} +U_{mm}
\end{eqnarray}
\noindent
where $U_{sv}$ is the energy $N$ non-interacting singly quantized vortices, 
$U_{vv}$ is the vortex-vortex interaction, $U_{mv}$ is the interaction 
energy between the FM and SC, and $U_{mm}$ is FM dot self interaction. 
In \cite{ours}, it was shown that the total energy of the system may be 
rewritten as follows:
\begin{eqnarray}
E=\int \left[\frac{n_s \hbar^2}{8m_e}\left(\nabla\varphi\right)^2
-\frac{n_s \hbar e}{4m_ec}\left(\nabla\varphi \cdot
{\bf{a}}\right) -\frac{1}{2} \bf{m}\cdot\bf{b}\right]d^2x
\label{eqn11}
\end{eqnarray}
\noindent
where $n_s$ is the two-dimensional superconducting electrons density and $m_e$
is their effective mass. $\hbar$ and $c$ are the Planck constant and the 
speed of light respectively. The vectorial quantities $\bf{a}$, $\bf{b}$ 
and ${\bf{m}}$ are the total vector potential and magnetic field due to 
the $N$ SC-vortices and the FM dot evaluated at the surface of the 
superconductor and the $2D$ magnetization of the FM texture. The phase 
gradient of the SC order parameter in the presence of $N$ vortices 
is ${\bf{\nabla}}\varphi=\sum_{n=1}^N \frac{({\bf{\rho}} -{\bf{\rho}}_n)\times 
\hat{z}}{|{\bf{\rho}} -{\bf{\rho}}_n|^2}$, where ${\bf{\rho_n}}$ 
is the location of the $n$-th vortex. In the presence of $N>1$ 
superconducting vortices the interaction of the vortices with the dot 
tries to lower the energy of the system due to its attractive nature 
while it is increased by the repulsive vortex-vortex interaction. If 
$N$ vortices are coupled to the FM dot then  we can recast the energy 
of the EMD-SC system using the identity 
$\int d^2 x \longrightarrow \frac{1}{4\pi^2}\int d^2 k$ as follows
\begin{eqnarray}
E &=& N \epsilon_0 ln(\frac{\lambda}{\xi}) + \epsilon_0 \lambda \sum_{i=1}^N 
\sum_{j\ne i}^N \int_0^\infty \frac{J_0 (\kappa|{\bf{\rho}}_i 
-{\bf{\rho}}_j|)}{1 +2\lambda \kappa} d\kappa \nonumber \\
& &-\frac{m_0 \phi_0 R_1 R_2}{\pi}\sum_{i=1}^N \Gamma(R_1,R_2,x_i,y_i) +E_{mm}
\label{eqn112}
\end{eqnarray}
\noindent
where $\kappa$ has a dimension of inverse length, 
$\epsilon_0=\frac{\phi_0^2}{16\pi^2 \lambda}$, and the function 
$\Gamma (R_1,R_2,x_i,y_i)$ is defined as follows
\begin{eqnarray} 
\Gamma(R_1,R_2,x_i,y_i)=\int \frac{ J_1 (G(k_x,k_y))
e^{i(k_x x_i + k_y y_i)}}{G(k_x,k_y)\left(1+2\lambda q\right)}d^2 q
\label{gamma}
\end{eqnarray}
\begin{figure}[ht]
\centering
  \includegraphics[angle=0,width=1.5in,totalheight=1.0in]{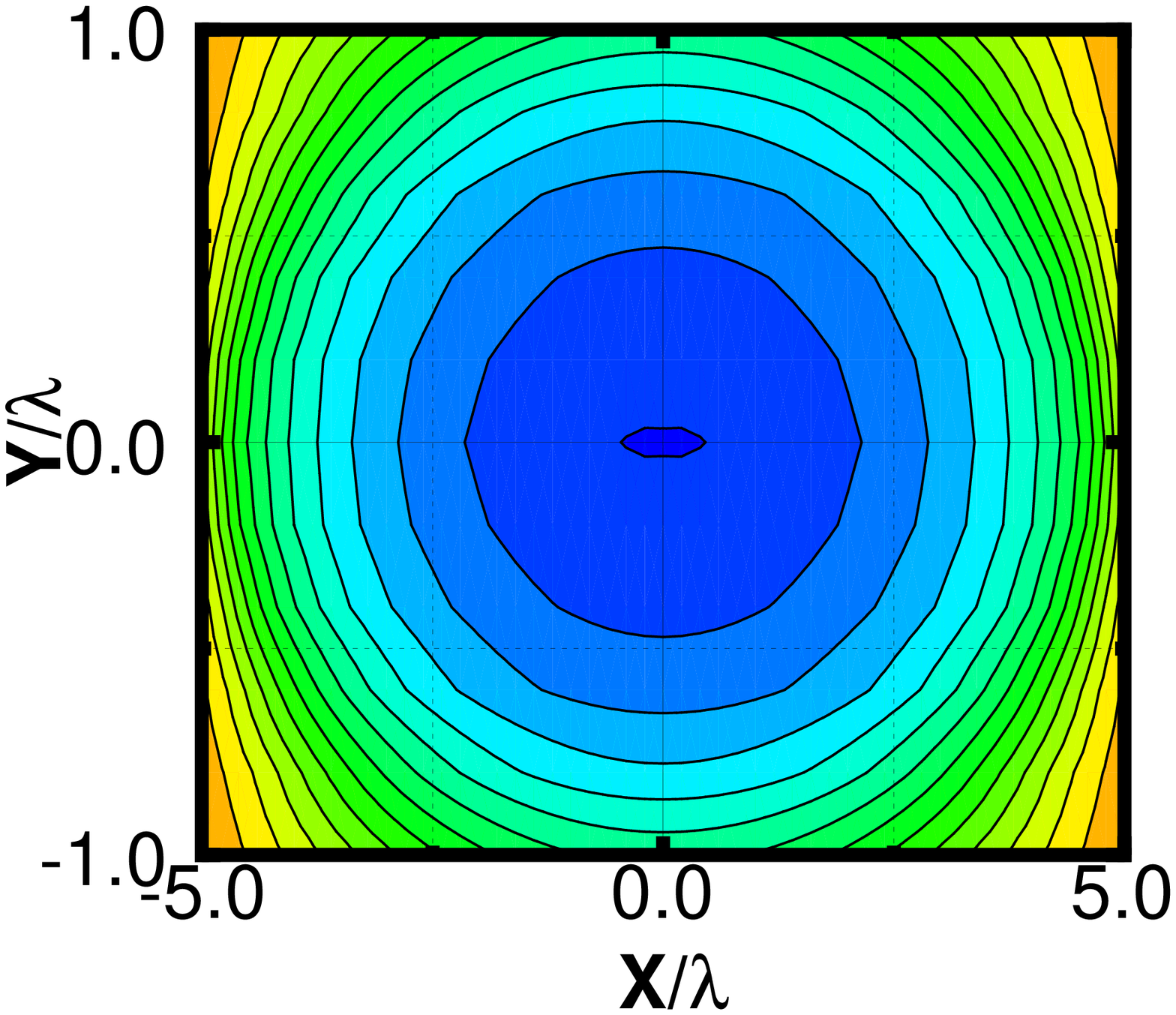}
  \hspace{0.1in}
  \includegraphics[angle=0,width=1.5in,totalheight=1.0in]{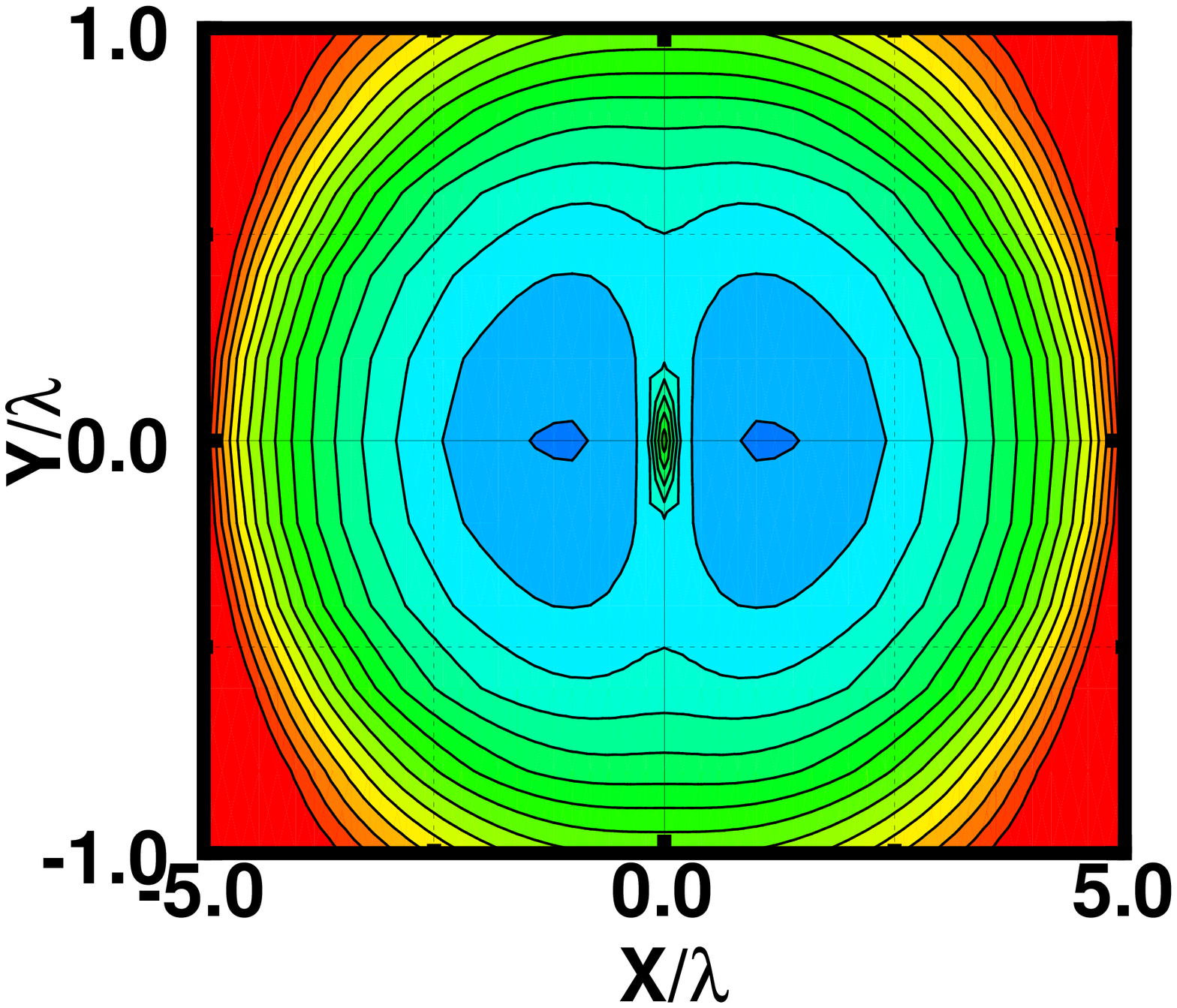}
  \caption{(Energy profiles for $N=1$ and $N=2$ in the 
    EMD-SC system. The EMD has major (minor) axis 
    $R_1=5\lambda$($R_2=\lambda$) and $m_0 \phi_0=10\epsilon_0$.}
\label{fig2}
\end{figure}

Vortex configurations for $N=1$ and $N=2$ are shown in Fig.(\ref{fig2}). 
For $N=1$, the vortex appears under the center of the dot while 
for $N=2$, vortices centers are located on the semi-major axis at equal 
distances from the center of the dot to minimize the total energy of 
the system. The degeneracy of the two vortices locations in the case of 
circular dot on the top of the SC film is lifted by the shape 
anisotropy of the dot elliptic dot. The creation of vortex configurations 
with $N>2$ requires larger values of $\delta_m =\frac{m_0 \phi_0}{\epsilon_0}$ 
to overcome the pearl energy and the repulsive vortex-vortex interaction. 
Vortex arrangements of $N>2$ depend on the ratio $\frac{R_2}{R_1}$. 
For $R_2 \sim \lambda$, vortices would line-up forming a straight chain of 
vortices extending under the semi-major axis of the dot. When 
$R_2 \gg \lambda$, the arrangement of vortices becomes more complex. Energy 
and vortex lattice structure for $N\gg 1$ can be found numerically 
by minimizing the total energy of the system given by Eq.(\ref{eqn112}).

\begin{figure}[ht]
  \centering
  \includegraphics[angle=0,width=3.0in,totalheight=2.0in]{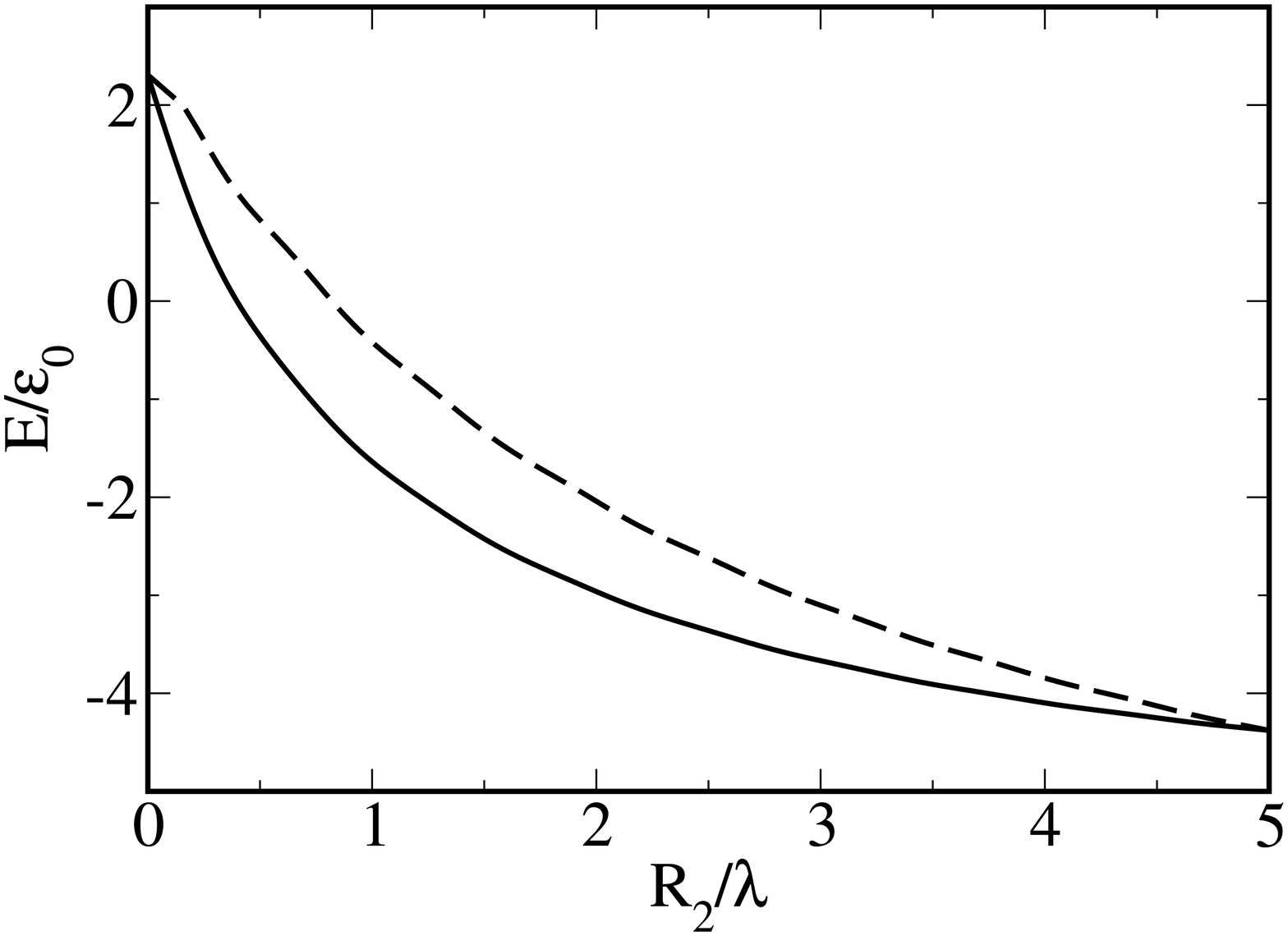}
  \caption{The solid line is energy of a single vortex in the presence of an 
    elliptic dot whose semi-major axis $R_1 =5\lambda $ as a function of $R_2$.
    The dashed line is the energy of a single vortex created by a circular 
    dot of radius $R_c =\sqrt{R_1 R_2}$. In either cases $\delta_m =2$ }
  \label{fig3}
\end{figure}

The energy of the single vortex depends on the eccentricity of the dot. To 
study this dependence, the energy of a single SC vortex coupled to an FM dot 
of fixed $R_1$ and variable $R_2$ must be calculated. The energy
dependence on $R_2$ is represented by the solid curve in Fig.(\ref{fig3}).
This shows that the lowest energy for $N=1$ configuration is reached when 
$R_2 =R_1$. However, this does not imply that spontaneous creation of 
superconducting vortices is more energy favorable if the dot is circular. 
This is because the magnetic flux supplied by the dot is maximum 
when $R_1 = R_2$. To better understand this, I compare the energy 
necessary to spontaneously create a single vortex by an elliptic dot with 
fixed $R_1$ and varying $R_2$ to the energy of a vortex created by a circular 
dot with the same per unit area magnetization $m_0$ and radius 
$R_c=\sqrt{R_1 R_2}$. The magnetic flux due to both dots is equal since their 
areas are equal. The curves in Fig(\ref{fig3}) shows that the creation of 
vortices by an elliptic dots is more energy favorable than those created by 
circular ones and has the same magnetic flux. The difference between the two 
curve is a reminisance of the shape anisotropy of the FM dot.   

\begin{figure}[ht]
  \centering
  \includegraphics[angle=0,width=3.0in,totalheight=2.0in]{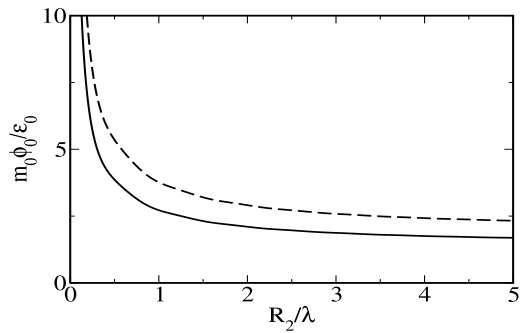}
  \caption{The solid (dashed) curve separate the regime without vortices 
    from the regimes with $N=1$ ($N=2$) vortices in the SC for $R_1 =5 
    \lambda$.}
\label{fig4}
\end{figure}

The appearance of a vortex under the dot changes the energy of 
the system by an amount of $\Delta= U_{sv} +U_{vv}+U_{mv}$ the vortex 
appear when $\Delta=0$. This criterion produces a surface in $3D$ space 
parametrized by $\frac{R_1}{\lambda}$, $\frac{R_2}{\lambda}$.
The surface $\Delta=0$ separates between regimes with and without vortices. 
Phase transitions from $N=0$ regime to $N=1$ and $N=2$ regimes are shown in 
Fig.(\ref{fig4}). Note that for strongly eccentric dot {\it{i.e.}} 
$R_2\ll R_1$ the spontaneous creation of vortices requires large values 
of $\frac{m_0 \phi_0}{\epsilon_0}$ due to small stray field of the dot.

Now, let us consider a square array of identical elliptic FM dots on the top 
of a superconducting thin film. Let all dots have their semi-major 
axis aligned along the $x$-axis, and they are well separated so that 
the dipolar interaction between them could be ignored. If $\delta_m$ is larger 
than a critical value then vortices appear under the dots. Due to the 
conservation of topological charge, equal number of antivortices will appear 
in the regions between the dots. In the presence of the antivortices the 
total energy of the system must include their interaction with the dot array 
and vortices and other antivortices in the system. For large enough 
array and a filling of one vortex per dot, vortices appear under the centers 
of the dots while antivortices will appear at the centers of the unit cells. 
This is so only if finite size effects are ignored. Since these effects 
violate the symmetry of the vortex lattice causing a shift in the locations 
of vortices and antivortices. Pinning forces acting on vortices are due to 
their interaction with the FM dots array and the vortex-antivortex 
interaction. Since the dots are well separated, the $i$-th vortex feels 
mostly the pinning potential created by the dot above it
\begin{eqnarray}
U_{mv}=-\frac{m_0 \phi_0 R_1 R_2}{\pi}\Gamma (R_1,R_2,x_i,y_i)
\label{eqnUmv}
\end{eqnarray}
The pinning by antivortices is isotropic and regular and can be represented by
a two-dimensional washboard potential. 
The pinning force exerted by the FM dot on a single vortex in the SC is 
derivable from $U_{mv}$ and its components are
\begin{eqnarray}
F_j (x_i,y_i)=-\frac{\imath m_0 \phi_0 R_1 R_2}{\pi} \Xi_j (R_1,R_2,x_i,y_i)
\label{pinforce}
\end{eqnarray}
\noindent
where $j=x,y$. Here the function $\Xi_j (R_1,R_2,x,y)$ is defined as follows:
\begin{eqnarray}
\Xi_j (R_1,R_2,x,y)=\int\frac{ k_j J_1 (G(k_x,k_y))
e^{i(k_x x + k_y y)}}{G(k_x,k_y)\left(1+2\lambda q\right)}d^2 q
\label{xifunc}
\end{eqnarray}

The shape anisotropy of the dot manifests itself in the pinning potential 
$U_{mv}$ and the pinning forces. Anisotropic pinning forces implies 
anisotropic transport properties such as anisotropic critical current. In 
other words the critical current $J_c$ for this system may depend on the angle 
$\theta$ between the driving current and the semi-major axis of the dots. 
It also must depend on $\delta_m$ and the eccentricity of the dots. 
For fixed value of $\delta_m$ and $\frac{R_2}{R_1}$, the strength of the 
transport anisotropy can be measured through the ratio 
$K_1 =\frac{J_c (\theta=\frac{\pi}{2})}{J_c (\theta=0)}$. To detect the 
effect of the dot's shape anisotropy on the transport properties of the 
underlaying superconductor, one can perform resistivity measurements while 
changing $\theta$. For a dots array whose dots are very eccentric, the 
measurements must reflect a decrease in the resistance of the sample as 
$\theta$ is increased down from 0 up to $\frac{\pi}{2}$. The full 
understanding of transport properties and the effect of the dot's shape 
anisotropy on vortex dynamics in this system is beyond the scope of this 
article.
 
In conclusion the system of a single elliptic dot on the top of a 
superconducting thin film is studied. The magnetic fields and screening 
currents for the FM-SC system are calculated self consistently using 
London-Maxwell electrodynamics. I showed that above some a threshold value 
of the dot's magnetization, the spontaneous creation of superconducting 
vortices becomes energy favorable. The appearance of the spontaneous vortex 
phase in a superconductor covered by an elliptic dot is more energy favorable 
compared to that created by a circular one that has the same area and 
magnetization per unit area. The phase transitions from regime without 
vortices to regimes with one and more vortices are studied. We also studied 
the pinning forces and transport properties for a superconducting thin film 
covered by an array of elliptic dots. The critical current for this system is 
affected by the shape anisotropy of the dots and it depend on the angle 
between the direction of the driving current and the semi-major axis of 
the dot. 

We would like to thank V. L. Pokrovsky, W. M. Saslow, and D. G. Naugle for 
helpful and fruitful discussions. This work was supported 
by NSF grants DMR 0103455 and DMR 0072115, DOE grant DE-FG03-96ER45598, and 
Telecommunication and Information Task Force at Texas A\&M University.


\begin{references}
\bibitem{schull} J. I. Martin, M. Velez, J. Nogues, and I. K. Schuller
Phys. Rev. Lett. {\bf 79}, 1929, (1997).

\bibitem{ketter} D. J. Morgan and J. B. Ketterson, Phys. Rev. Lett. {\bf 80},
3614(1998).

\bibitem{mosch} Van Bael MJ, Bekaert J, Temst K, Van Look L,
Moshchalkov VV, Bruynseraede Y, Howells GD,
Grigorenko AN, Bending SJ,  Borghs G
Phys. Rev. Lett. {\bf 86}, 155 (2001);
%
 Lange M, Van Bael MJ, Van Look L, Temst K, Swerts J,
Guntherodt G, Moshchalkov VV, Bruynseraede Y Europhys. Lett.  {\bf
51}, 110 (2001);
%
Bending SJ, Howells GD, Grigorenko AN, Van Bael MJ,
Bekaert J, Temst K, Van Look L, Moshchalkov VV, Bruynseraede Y,
Borghs G, Humphreys RG
Physica C{\bf 332}, 20 (2000);

\bibitem{bula} L. N. Bulaevskii, A. I. Buzdin, M. L. Kulic and S. V. Panyukov,
Adv.Phys. {\bf 34}, 175 (1985).

\bibitem{PL1} I. F. Lyuksyutov and V. L. Pokrovsky Phys. Rev. Lett. {\bf 81},
2344 (1998).

\bibitem{PL2} I. F. Lyuksyutov and V. L. Pokrovsky,
{\it{Magnetism Controlled Vortex Matter.}} cond-mat/9903312.

\bibitem{ours} S. Erdin, M. A. Kayali, I. F. Lyuksyutov and V. L. Pokrovsky,
Phys. Rev. B {\bf 66}, 014414 (2002).

\bibitem{peeters} I. K. Marmorkos, A. Matulis and F. M. Peeters Phys. Rev. B
{\bf 53}, 2677 (1996).

\bibitem{mine} M. A. Kayali, Phys. Lett. A {\bf 298}, 432 (2002).

\bibitem{milosovic} M. V. Milosevic and F. M. Peeters, {\it{The
interaction between a superconducting vortex and an out-of-plane
magnetized ferromagnetic disk: influence of the magnet geometry.}}
cond-mat/0303310.

\bibitem{erdin} S. Erdin, I. Lyuksyutov, V. Pokrovsky and V. Vinukor, Phys. 
Rev. Lett.{\bf{80}} (2002).

\bibitem{sasik} R. Sasik, T. Hwa, {\it{Enhanced Pinning of Vortices in
Thin Film Superconductors by Magnetic Dot Arrays.}}
cond-mat/0003462.

\bibitem{KP} M. A. Kayali and V. L. Pokrovsky, {\it{Anisotropic 
Transport Properties of Ferromagnetic-Superconducting Bilayers.}}, 
cond-mat/030383873.

\bibitem{buzdin1} A. Buzdin, M. Daumens, Physica C {\bf 294}, 257 (1998).

\bibitem{buzdin2} A. Buzdin, D. Feinberg, Physica C {\bf 256}, 303 (1996).


\bibitem{buzdin4} A. Buzdin, M. Daumens, Physica C {\bf 332}, 108 (2000).

\bibitem{buzdin5} C. Meyers, A. Buzdin, Phys. Rev. B {\bf 62}, 9762 (2000).


\bibitem{abrik} A. A. Abrikosov, {\it{Introduction to the Theory
of Metals}} (North-Holland, Amsterdam, 1986).


\end{references}
\end{document}